# Human resources management system for Higher Education institutions

## Ivona Zakarija

Department of Electrical Engineering and Computing
University of Dubrovnik
Ćira Carića 4, Dubrovnik, Croatia
Phone: +358 (0)20-445 742  E-mail: ivona.zakarija@unidu.hr

## Zoran Skočir

Faculty of Electrical Engineering and Computing
University of Zagreb
Unska 3, Zagreb, Croatia
Phone: +358 (0)1- 6129 831  E-mail: zoran.skocir@fer.hr

## Krunoslav Žubrinić

Department of Electrical Engineering and Computing
University of Dubrovnik
Ćira Carića 4, Dubrovnik, Croatia
Phone: +358 (0)20-445 742  E-mail: krunoslav.zubrinic@unidu.hr

**Abstract - In an environment where employees and their experience are of central value to the company, human resources management (HRM) represents a significant aspect of business efficiency. In this paper we present a model of an HRM information system for universities. Special attention is paid to aspects of the system that support processes specific to science and higher education. UML was used for modelling the global and detailed system architecture and database model. FURPS+ methodology was used for classification of requirements and the MoSCoW method for analysis of requirement priority.**

## 1. INTRODUCTION

A human resources management (HRM) system is an important part of every business organization. Science and higher education institutions are no different in this respect. In addition to regular human resource (HR) data, as in any other organization, such institutions need to include specific data in accord with national and institutional laws and regulations. As a result it may be necessary for such institutions to develop a specialised HRM system.

In this paper we describe the development of the HRM system in use at the University of Dubrovnik .

The basis for the development of a model for the HRM system is described in the second section of the paper. The third section describes the methodology used for the requirement analysis. The fourth section explains in detail the academic professional development structure in Croatian science and education institutions, which determines the principal part of a specific HR system for science and education institutions.

The structure of the information system is described in the fifth section. Special attention is paid to processes specific to science and higher education. The sixth section describes the architecture of the final HRM system implemented at the University of Dubrovnik.

## 2. BASIS FOR MODEL CREATION

Interviews with the head of Human Resources at the University of Dubrovnik as well as a study of their business data served as the main sources of information used for the development of the model. Research on the Law on Higher Education and Science provided additional information for setting up a model.

For the first stage we implemented a version of the HRM system as a character application, applicable to a range of industrial and business situations with a view to establishing the feasibility of database re-engineering and migration to another platform. Users entered real data in order to verify that the application suited their needs. After a series of test iterations we began the development of a new version of the HRM application.

At the next stage we re-wrote a complete legacy application on Windows GUI platform. In order to ensure parallel functioning of the old and newer versions of the HRM system we were forced to keep the legacy data structure but we conducted a database re-engineering wherever possible. For that reason there is some redundancy in the database.

The final development stage involved customisation by incorporating some new functionalities and features specific to higher educational institutions according to customer requirements.

### 3.   REQUIREMENT ANALYSIS

The requirements engineering approach was used to capture system requirements. *Requirements engineering* [1] is a term used to describe the activities involved in eliciting, prioritizing, documenting, and maintaining a set of requirements for a software system. It is a process of assessment of what stakeholders need the system do for them.

Since there are often conflicting requirements that must be balanced, a degree of compromise and negotiation is generally required. According to FURPS + classification requirements are placed into these categories:

- functional requirements describing the functionality that the system is to execute; what behaviour the system should offer (capabilities).
- non-functional requirements are ones that act to constrain the solution

FURPS+ methodology has been adopted by many organizations and integrated into an international standard ISO/ IEC 9126. [2] It is an acronym for a model for classifying software quality attributes:

- Functionality
- Usability
- Reliability
- Performance
- Supportability

"+" (ancillary) :

- Implementation
- Interfaces
- Operations
- Packaging
- Licensing

We used the MoSCoW analysis to achieve a clear prioritization of requirements. MoSCoW stands for:

- Must Have
- Should Have
- Could Have
- Won't Have This Time Around [3]

Requirements are documented in the standard form where functional requirements are separated and the MoSCoW list is integrated. Each requirement has a *Priority* attribute that can take one of the values M, S, C, or W. Modelling customer requirements in this way make a significant contribution in determining the project scope.

### 4.   ACADEMIC PROFESSIONAL DEVELOPMENT

University personnel comprise two major groups:

- administrative staff
- academic staff

Besides existing standard personnel data the HRM system for Higher Education institutions must take account of academic professional development. Under the law    [4] in higher institutions scientific research activities can be performed by:

- scientists and researchers
- teachers and associates

of appropriate level. Employees occupy academic positions according to [5] with official document issued by higher institutions.

#### 4.1   Scientists and Researchers

The work of scientific research is carried out by qualified scientists and researchers, who, under existing rules and procedures[6][4][7], can be classified into the following grades [8]:

Scientists:

- research associates, senior research associates and research advisors,

Researchers:

- expert assistant, younger assistent, assistent, senior assistent

#### 4.2   Register of Researchers

The Ministry of Science, Education and Sports keeps records of scientists and researchers in the Register of Researchers. The following categories appear in the Register:

- research associates, senior research associates and research advisors,
- assistant professors, associate professors and full professors.
- external associates – assistant professors and senior assistant professors,
- persons with a doctoral degree.

Universities initiate procedures of entry into the Register based on the submission of an application to the Ministry with all relevant documentation. Each person has a unique ID number assigned – identity number of scientists. In the HRM system applications are supported by additional employee data.

#### 4.3   Teachers and Associates

The work of education is carried out by either full time teachers or associate staff; most university teachers are also active as scientists and researchers at their university or other research institute. Under the rules and procedures in force teachers can be of the following grades:

- scientific-research grades:
  - assistant professor, associate professor, full professor, professor emeritus
- teaching grades:
  - lecturer, senior lecturer, professor of high school, lector, senior lector, repetiteur, senior repetiteur

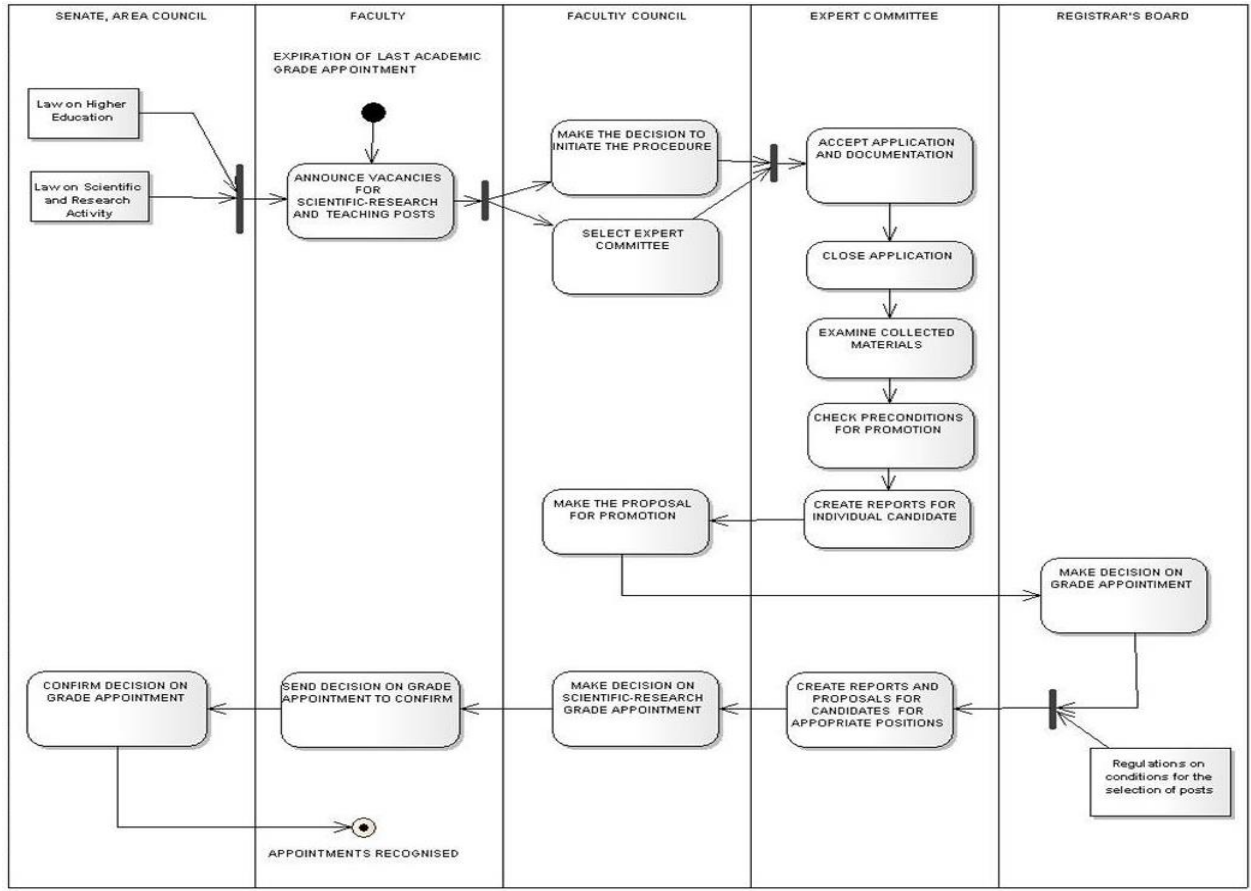

**Fig. 1 UML activitiy diagram shows conducting of the grade appointment procedure**

Associate grades:
  - expert assistant, younger assistent, assistent, high school assistant, senior assistent

Academic grades with their corresponding required skills in teaching and research are defined by the University Law. Promotion of academic staff is designed to recognise and reward sustained excellence. Assessments are made by committees of peers through a process designed to enable fair and consistent application of standards. To be promoted, academic staff members must, on objective evidence, attain an appropriate standard.

## 5. DESCRIPTION OF INFORMATION SYSTEM MODEL

Based on the results of research and interviews a information system model has been set up. HRM system consists of following segments as presented in Figure 2. We will not in detail elaborate each particular segment, though special attention is paid to aspects of the system that support processes specific for science and higher education.

### 5.1 Conducting the grade appointment procedure

Universities teachers are appointed to scientific-research and teaching grades for a period of five years.

Appointments are made on the basis of open application. Based on the regulations and procedures in force [9] the University announces vacancies for scientific-research and teaching post appointments. The University aims to ensure equity, transparency and fairness in all aspect of the appointment process.

The University must inform all applicants of the criteria and procedures for appointment. Faculty Council make a decision to initiate the procedure and select the board members of the Expert Committee.

After the official announcement the Committee accept applications and documentation and when the application is closed examines those materials. Based on their reports the competent Registrar Board make a decision and list applicants for promotion of grade and academic position. After confirmation from Area Council/Senate the appointments are recognised. Figure 1 shows the UML activity diagram of conducting the grade appointment procedure.

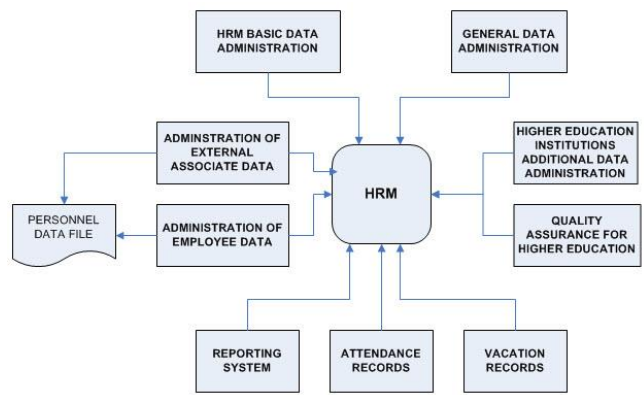

**Fig. 2 HRM system segments**

The HRM system supports appointment/re-appointment processes through the module The academic grade records as shown on Figure 4. It is a part of Higher Education additional data administration segment. Detailed data on all recognized grade appointments are recorded, and the system gives a review of warnings and notifications of grade appointment expiry.

This is important because the procedure for a grade application must be initiated 3 months before the grade expires and the institution must prepare applications within that period of time.

Evidence for assessment for promotion to a higher academic grade may include: articles in journals, including electronic journals; articles in the proceedings of, and presentations to, national and international conferences. In the HRM system for employees records of such scientific papers are in the form of title, type of work, publishing date, and URL to the CROSBI where it is stored in electronic form. A module of HRM system is called Published scientific papers records. Universities can take an up to date bibliography for employees without the need to enter a large amount of data. The Croatian Scientific Bibliography (CROSBI) is a digital archive which stores all scientific publications in Croatia [10].

Complete documentation related to promotion procedure applications, committee structure, decision of grade appointment and decisions of Registrar Board can be stored electronically in a document repository. In the HRM system it is called Attached document records – and involves simply entering the path where a document is stored in the repository. ActiveX control built into the HRM system ensures access to documents in the repository regardless of format (.pdf ,. doc,. xls,. jpg,. html)

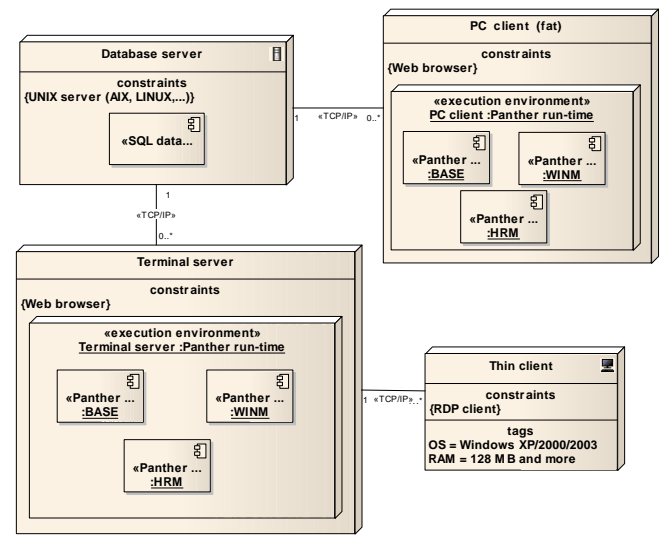

**Fig. 3 System architecture**

## 6. SYSTEM ARCHITECTURE

To achieve easier implementation and maintenance of the application two-tier architecture is employed in our system. The architecture is composed of two layers: the database layer and the business logic. The database layer and the business logic are located on separate physical machines.

There are two kinds of clients:

«thin clients» - connect via RDP protocol to the application terminal server where the complete application logic is performed. Terminal server handles the database and the business logic. The main role of «thin client» is to present application data in user interface.

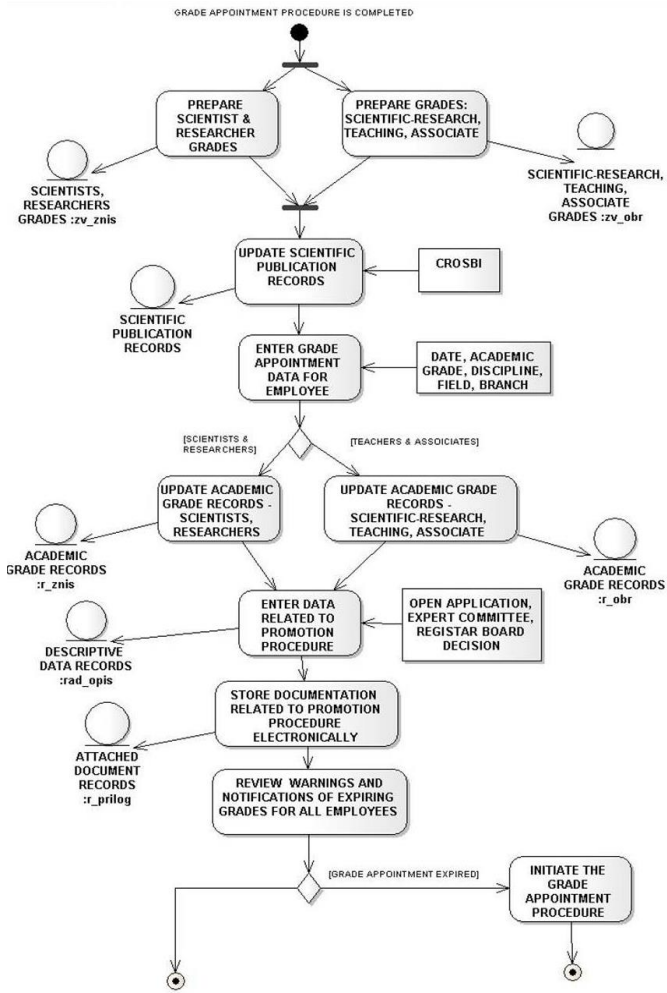

**Fig. 4 UML activity diagram showing a module for managing the academic grade records**

«fat clients» - where the complete application logic is performed; such clients connect directly to the database server. Figure 3 shows system architecture.

## 7. CONCLUSION

In the paper an HRM system for higher education institutions is presented. Implementation of this system at the University of Dubrovnik has confirmed that it provides effective support for certain specific processes in the appropriate context. The use of the system for grade appointment procedures is shown in detail.